\RequirePackage{fix-cm}
\RequirePackage{amsmath}
\documentclass[twocolumn]{svjour3}          % twocolumn
\smartqed  % flush right qed marks, e.g. at end of proof
\usepackage{graphicx}
\usepackage{glossaries}
\usepackage{siunitx}
\usepackage{pgf,pgffor}
\usepackage[colorlinks=false]{hyperref}
\usepackage[misc]{ifsym}

\begin{document}

\title{Reducing the dependence of the neural network function to systematic uncertainties in the input space}
%\subtitle{}/

%\titlerunning{Explaining the Performance of Neural Networks}

\author{Stefan Wunsch \and Simon J\"orger \and Roger Wolf \and G\"unter Quast}


\institute{
Stefan Wunsch\textsuperscript{1,2} (corresponding author) \at
stefan.wunsch@cern.ch
\and
Simon J\"orger\textsuperscript{1} \at
simon.joerger@cern.ch
\and
Roger Wolf\textsuperscript{1} \at
roger.wolf@cern.ch
\and
G\"unter Quast\textsuperscript{1} \at
guenter.quast@kit.edu
\and
\textsuperscript{1} Karlsruhe Institute of Technology, Institute of Experimental Particle Physics, Karlsruhe, Germany\\
\textsuperscript{2} CERN, Geneva, Switzerland
}


\maketitle

\begin{abstract}

Applications of neural networks to data analyses in natural sciences are complicated by the fact that many inputs are subject to systematic uncertainties. To control the dependence of the neural network function to variations of the input space within these systematic uncertainties, several methods have been proposed. In this work, we propose a new approach of training the neural network by introducing penalties on the variation of the neural network output directly in the loss function. This is achieved at the cost of only a small number of additional hyperparameters. It can also be pursued by treating all systematic variations in the form of sta\-tis\-ti\-cal weights. The proposed method is demonstrated with a simple example, based on pseudo-ex\-pe\-ri\-ments, and by a more complex example from high-energy particle physics.

%\keywords{Neural Networks \and Systematic Uncertainties \and High-Energy Particle Physics}
\end{abstract}

\glsresetall

\section{Introduction}
\label{sec:Introduction}

Neural network (NN) techniques are in wide and increasing use to solve classification and regression tasks in the analysis of high-energy particle physics data. Examples of their use in physics object identification, e.g. at the LHC experiments ATLAS and CMS, are the classification of particle jets induced by heavy flavor quarks~\cite{Aad:2015ydr,Sirunyan:2017ezt} and the identification of $\tau$ leptons~\cite{Aad:2015unr,Sirunyan:2018pgf}. Examples for data analyses that make use of NNs not only for object identification, but to distinguish between signal- and background-like samples are the latest analyses of Higgs boson events in association with third generation fermions, at the LHC~\cite{Aaboud:2018urx,Aaboud:2018zhk,Sirunyan:2018hoz,Sirunyan:2018kst,CMS-PAS-HIG-18-032}. These classification tasks usually aim at the distinction of a signal from one or more background processes. They are characterized by a relatively small number of input parameters to the NN, of one or two orders of magnitude, which may reveal non-trivial correlations among each other.

Each physics measurement is subject to systematic uncertainties, which have to be propagated from the input space $\mathbf{x}=\{x_{i}\}$ to the NN output $f(\mathbf{x})$. This usually happens in terms of variations of a given input parameter $x_{i}$ within its uncertainties $\Delta_{i}$. We abbreviate the set of $\Delta_{i}$ by $\mathbf{\Delta}=\{\Delta_{i}\}$ and the set of modified input parameters by $\mathbf{x+\Delta}=\{x_{i}+\Delta_{i}\}$. These variations may be implemented in the form of variations of the actual values of $x_{i}$, or such that a sample, with a given value of $x_{i}$, enters the analysis with a different statistical weight, also referred to as reweighting throughout this text. Unlike varying the values of $x_{i}$, reweighting does not rely on a reprocessing of the dataset and therefore generally implies significantly smaller computational costs.

The possibility to implement prior information about systematic uncertainties already in the NN training is motivated by two considerations: Firstly, a powerful distinction between classes in principle, can be considerably compromised by systematic uncertainties. Integrating prior knowledge of uncertainties in the NN training helps in guiding the NN to focus on features in the input space that are less prone to such a performance degradation. This may even result in a gain for the analysis performance, as observed in Ref.~\cite{shimmin2017decorrelated}.
Secondly, the dependence of a systematic variation of a given feature $x_{i}$ on other parameters $\{x_{j}\}, j\neq i$ in the input space, might only be poorly known, or even unknown, and the user might want to generally uncorrelate the NN output from this uncertainty to assure a reliable response of the NN to the given task. Both points raise interest in training the NN with the boundary condition that the dependence of $f(\mathbf{x+\Delta})$ on $\mathbf{\Delta}$ should be minimal.

One way of achieving this decorrelation of $f(\mathbf{x+\Delta})$ from $\mathbf{\Delta}$ that has been proposed in the past and that we will refer to in more detail throughout this paper, makes use of a secondary NN that is trained in addition to the primary NN in an iterative procedure, resulting in an adversarial architecture~\cite{Goodfellow:2014upx} for robust binary classification~\cite{louppe2017learning}. This secondary NN has the task of drawing information of the systematic variation from the output of the primary NN. The output of the secondary NN is then included in the loss function of the primary NN as part of a minimax optimization problem. The resulting setup becomes insensitive to the systematic variation of the inputs. This method requires a relatively complex iterative training procedure; it introduces a large and to some extent arbitrary number of new hyperparameters implied by the choice of the architecture of the secondary NN, and requires the resampling of $x_{i}$ within its uncertainties $\Delta_{i}$.

Another approach to decorrelate $f(\mathbf{x+\Delta})$ from $\mathbf{\Delta}$ is to include the knowledge about systematic uncertainties in a systematics-aware objective function as proposed in Refs.~\cite{deCastro:2018mgh} and~\cite{Charnock:2018ogm}. An approach related to boosted decision trees is implemented by splitting the tree nodes using the signal significance including systematic uncertainties as objective, resulting in a classifier that successfully reduces the impact of systematic uncertainties on the result~\cite{Xia:2018kgd}. A similar approach for NNs has been studied in Ref.~\cite{Elwood:2018qsr}. A comparison of systematics-aware learning techniques in high-energy particle physics has been carried out in Ref.~\cite{estrade:hal-01715155}. In addition to the adversarial approach discussed above this study includes a comparison to data perturbation and augmentation, and tangent propagation~\cite{NIPS1991_536}.

In our approach we implement a penalty on the differences between the NN output obtained from the nominal value of $x_{i}$ and its variations $\Delta_{i}$, directly into the loss function. For this purpose we use histograms of $f(\mathbf{x})$ and $f(\mathbf{x+\Delta})$ filled during each training batch. The number $n_{k}$ of histogram bins $\{k\}$, and the batch size $n_{b}$ are hyperparameters of the training. To guarantee a differentiable loss function for the optimization of the trainable parameters of the NN, the histogram bins are blurred by a filter function applied to each sample $b$ of the training batch, affected by the uncertainty variations, where $b$ corresponds to a single sample represented by a point associated to each respective training dataset in the input space $\mathbf{x}$. We use Gaussian functions $\mathcal{G}_{k}(\mathbf{x})$, normalized to $\max\left(\mathcal{G}_{k}(\mathbf{x})\right)=1$ as filters, where the mean and standard deviation are given by the center and half-width of histogram bin $k$. The count estimate can then be written as $\mathcal{N}_{k}(f(\mathbf{x})) = \sum_{b}\mathcal{G}_{k}(f(x_{b}))$, and the loss function consists of the two parts
\begin{align*}
    &L_{\Lambda} =  L^{\prime} + \lambda\,\Lambda(\mathbf{x}, \mathbf{\Delta}) %\\
    \vphantom{\frac{1}{n_{k}}\sum_{k}{\Bigl(\frac{\mathcal{N}_{k}(f(\mathbf{x})) - \mathcal{N}_{k}(f(\mathbf{x+\Delta}))}{\mathcal{N}_{k}(f(\mathbf{x}))}\Bigr)^{2}}}\\
%\end{align*}
%with:
%\begin{align*}
    &\text{with:} \\
    &\Lambda(\mathbf{x},\mathbf{\Delta}) = \frac{1}{n_{k}}\sum_{k}{\Bigl(\frac{\mathcal{N}_{k}(f(\mathbf{x})) - \mathcal{N}_{k}(f(\mathbf{x+\Delta}))}{\mathcal{N}_{k}(f(\mathbf{x}))}\Bigr)^{2}},
\end{align*}
where $L^{\prime}$ corresponds to the loss function of the primary task, like for example the cross-entropy function for a classification task, and $\Lambda(\mathbf{x}, \mathbf{\Delta})$ to the term that penalizes differences in the NN function between $f(\mathbf{x})$ and $f(\mathbf{x+\Delta})$. The factor $\lambda$ controls the influence of the penalty and adds another hyperparameter to the training. The count estimate $\mathcal{N}_{k}(f(\mathbf{x+\Delta}))$ can be derived from $\mathcal{N}_{k}(f(\mathbf{x}))$ in terms of reweighting, such that no reprocessing of the dataset during the training procedure is required.

In this approach more than one uncorrelated uncertainty simply adds to the sum of $\Lambda_{\ell}(\mathbf{x}, \mathbf{\Delta})$, for $\ell$ uncorrelated uncertainties. Two fully (anti-) correlated uncertainties should be represented by a common variation for both uncertainties at the same time. While an exact modeling of correlations across uncertainties may not always be exactly known this knowledge is not strictly required by the method, as long as the loss function converges to its minimum and solves the defined task. The parameters $\lambda_{\ell}$ correspond to further hyperparameters, whose values relative to each other define different tasks of the NN training. We would like to emphasize that the use of a histogram of $f(\mathbf{x})$ (and $f(\mathbf{x+\Delta})$ respectively) in the loss function might lead to a suboptimal performance with respect to the direct use of $f(\mathbf{x})$. Also we do not claim the resulting discriminator to be optimal for the final measurement.

In Section~\ref{sec:Application_to_a_toy} we demonstrate the method on a simple example based on pseudo-experiments. A more complex analysis task typical for high-energy particle physics is studied in Section~\ref{sec:Application_to_a_more_complex}. We summarize our findings in Section~\ref{sec:Summary}.

\section{Application to a simple example based on pseudo-experiments}
\label{sec:Application_to_a_toy}

To illustrate our approach, we refer to a simple example based on pseudo-experiments that has also been used in Ref.~\cite{louppe2017learning}. It consists of two variables $x_{1}$ and $x_{2}$, which are the input to separate two classes, in the following labeled as signal and background. The input space is visualized in Fig.~\ref{fig:toy1_data}. A systematic uncertainty for the background class is introduced by two variations of $x_{2}$ by $\pm1$. We consider only the discrete variations that quantify the difference between $\mathcal{N}_{k}(f(\mathbf{x}))$ and $\mathcal{N}_{k}(f(\mathbf{x+\Delta}))$, which is sufficient to define the part of $\Lambda(\mathbf{x},\mathbf{\Delta})$ to be minimized during the training process. We have checked that a Gaussian sampling with a standard deviation of $\sigma=1$, as applied in~\cite{louppe2017learning} would lead to the same result in a more complex setup.

\newsavebox{\meanSig}
\savebox{\meanSig}{$\left(\begin{smallmatrix}0 & 0\end{smallmatrix}\right)$}
\newsavebox{\meanBkg}
\savebox{\meanBkg}{$\left(\begin{smallmatrix}1 & 1\end{smallmatrix}\right)$}
\newsavebox{\stdSig}
\savebox{\stdSig}{$\left(\begin{smallmatrix}\hphantom{-}1\hphantom{.0}&-0.5\\-0.5&\hphantom{-}1\hphantom{.0}\end{smallmatrix}\right)$}
\newsavebox{\stdBkg}
\savebox{\stdBkg}{$\left(\begin{smallmatrix}\hphantom{-}1\hphantom{-}&\hphantom{-}0\hphantom{-}\\\hphantom{-}0\hphantom{-}&\hphantom{-}1\hphantom{-}\end{smallmatrix}\right)$}
\begin{figure}
\centering
\includegraphics[width=0.8\linewidth]{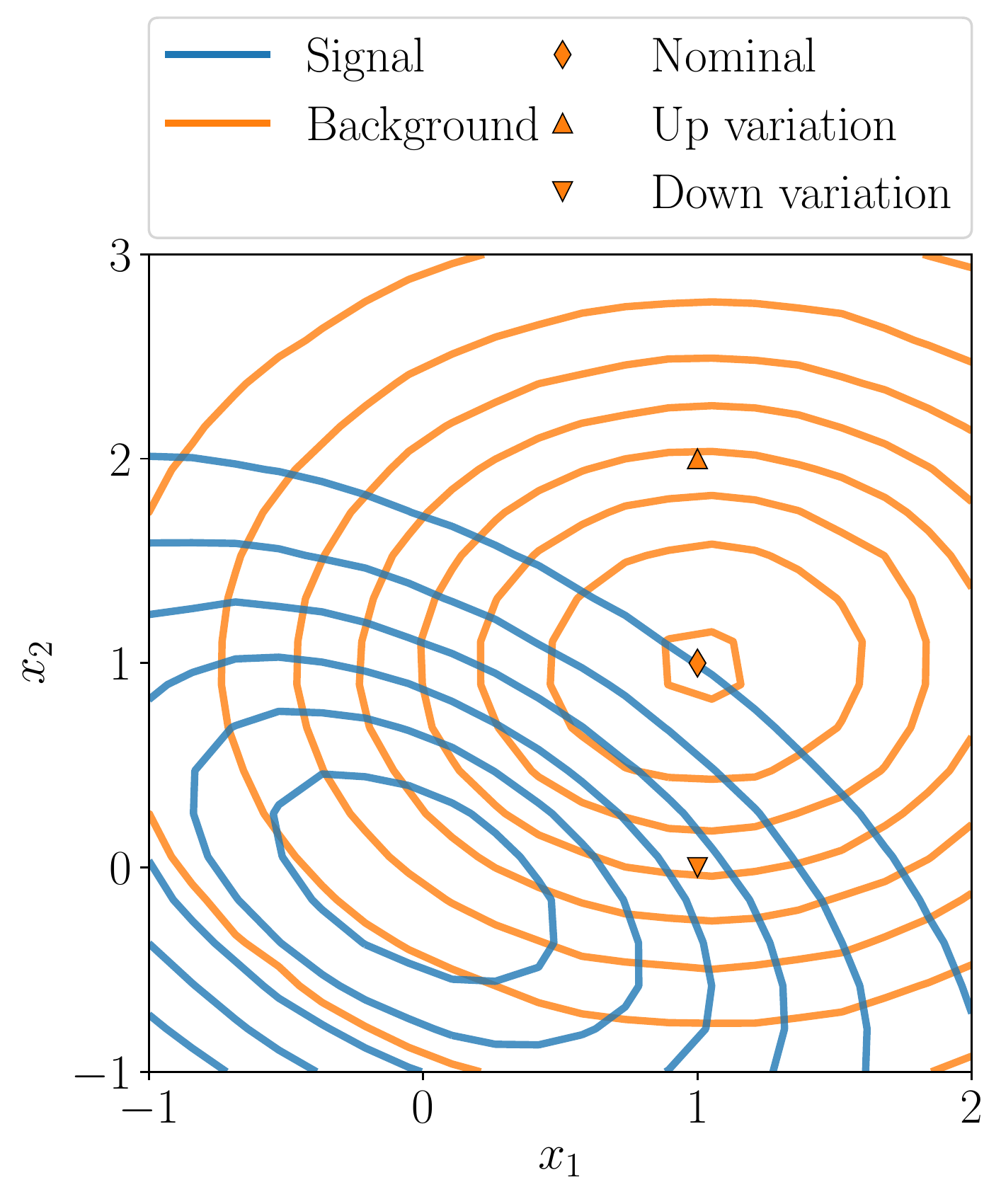}
\caption{Distribution of the input variables in the example of two classes labeled as signal and background, given in Section~\ref{sec:Application_to_a_toy}. Two multivariate Gaussian distributions are centered around \usebox{\meanSig} and \usebox{\meanBkg} with the covariance matrices \usebox{\stdSig} and \usebox{\stdBkg}, respectively. An additional uncertainty may lead to variations of the mean of the background sample on the y-axis as indicated for the mean values of the background distribution in the figure.}
\label{fig:toy1_data}
\end{figure}

\begin{figure*}
\centering
\includegraphics[width=0.4\linewidth]{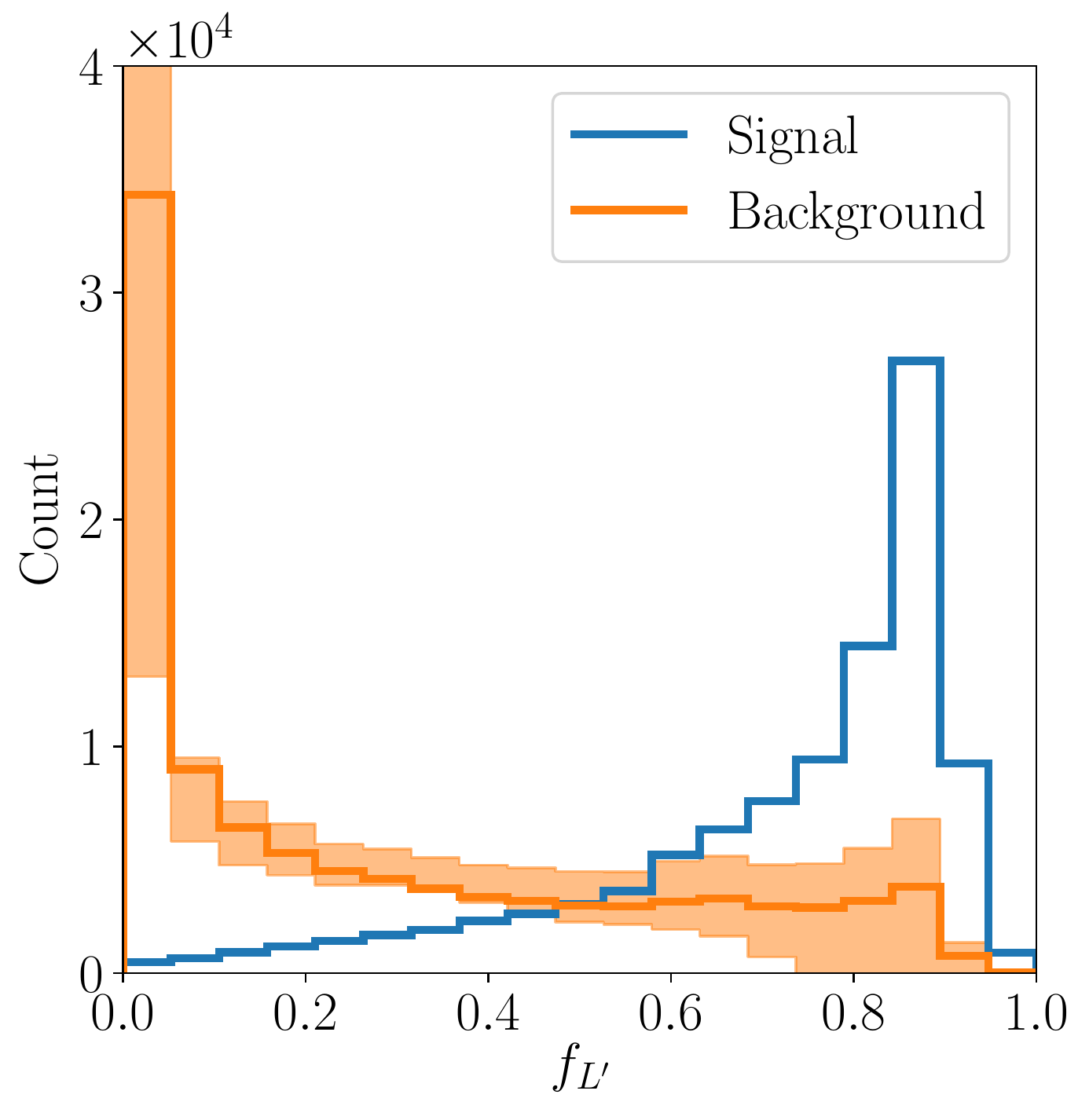}%
\includegraphics[width=0.4\linewidth]{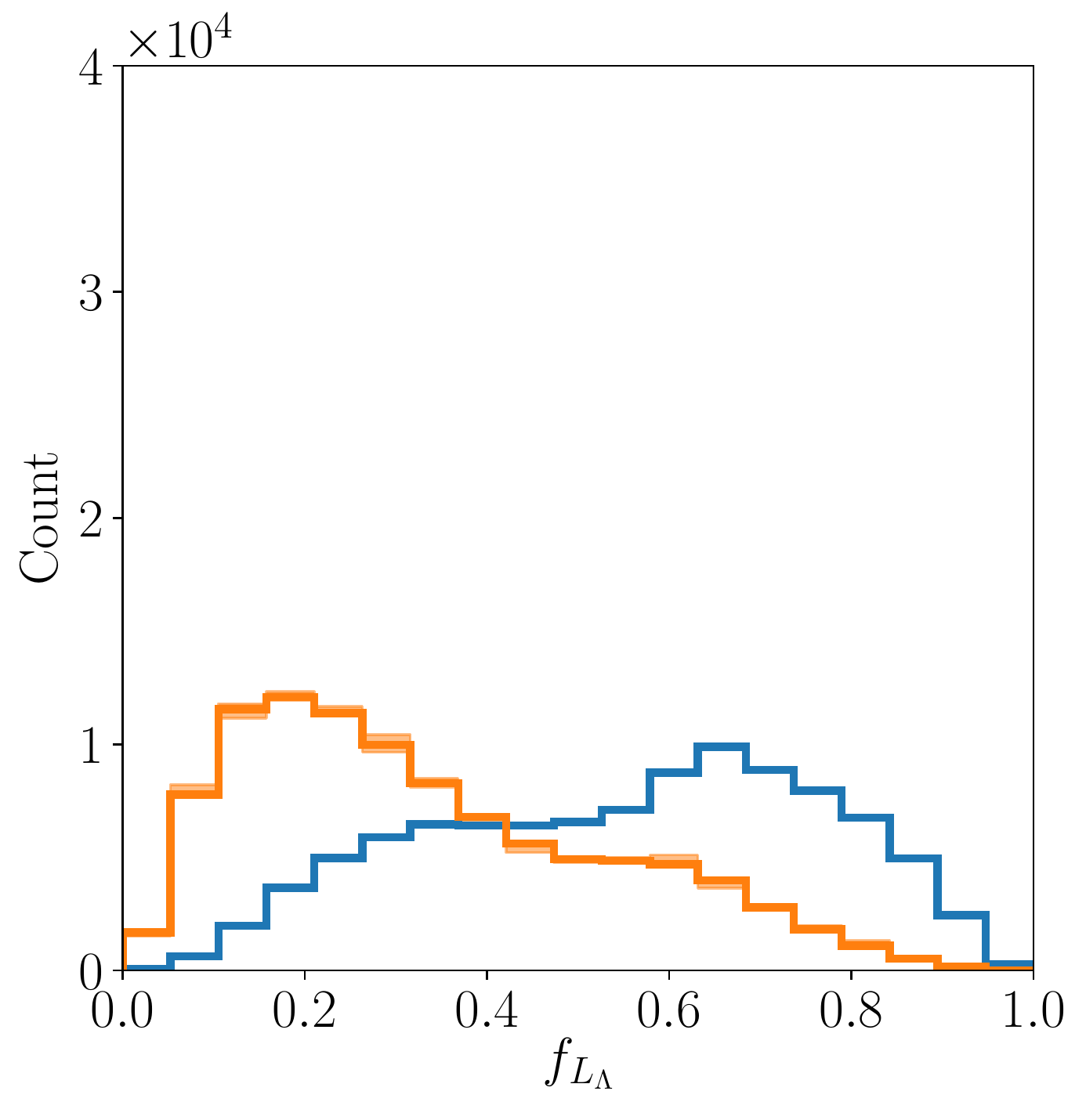}
\caption{Distribution of the NN output for a classifier trained (left) with a cross-entropy function only ($f_{L^{\prime}}$), and (right) with an additional term penalizing the variation of the NN output with the systematic variation of $x_{2}$ ($f_{L_{\Lambda}}$). The colored band around the distribution of the NN output of the background sample shows the effect of the systematic variation of $x_{2}\pm1$. Note that the uncertainty band in the first bin of the background histogram in the left subfigure is cut off.}
\label{fig:toy1_separation}
\end{figure*}

The NN used to solve the classification tasks consists of two hidden layers with 200 nodes each, with rectified linear units as activation functions~\cite{glorot2011deep} and a sigmoid activation function for the output layer. The trainable parameters are initialized using the Glorot algorithm~\cite{glorot2010understanding}. The optimization is performed using the Adam algorithm~\cite{kingma2014adam} with a batch size of $\SI{e3}{}$. Our choice for $L^{\prime}$ is the cross-entropy function. For $\Lambda$, we use 10 equidistant bins in the range $[0, 1]$ of the NN output. We have not observed any significant performance dif\-fe\-ren\-ces by varying the number of histogram bins within reasonable boundaries, though. Finally we set $\lambda$ to 20. The training on $\SI{5e4}{}$ events is stopped if the loss obtained from the training dataset has not decreased for five epochs in sequence, on an independent validation dataset of the same size. In addition, we use $\SI{e5}{}$ events for testing and to produce the figures to illustrate the result. The impact of the systematic variations on the NN output is shown in Fig.~\ref{fig:toy1_separation} for the case of a classifier trained with a loss function given only by $L^{\prime}$ ($f_{L^{\prime}}$) and a classifier based on a loss function including the additional penalty term $\Lambda$ ($f_{L_{\Lambda}}$).

\begin{figure*}
\centering
\includegraphics[width=0.45\linewidth]{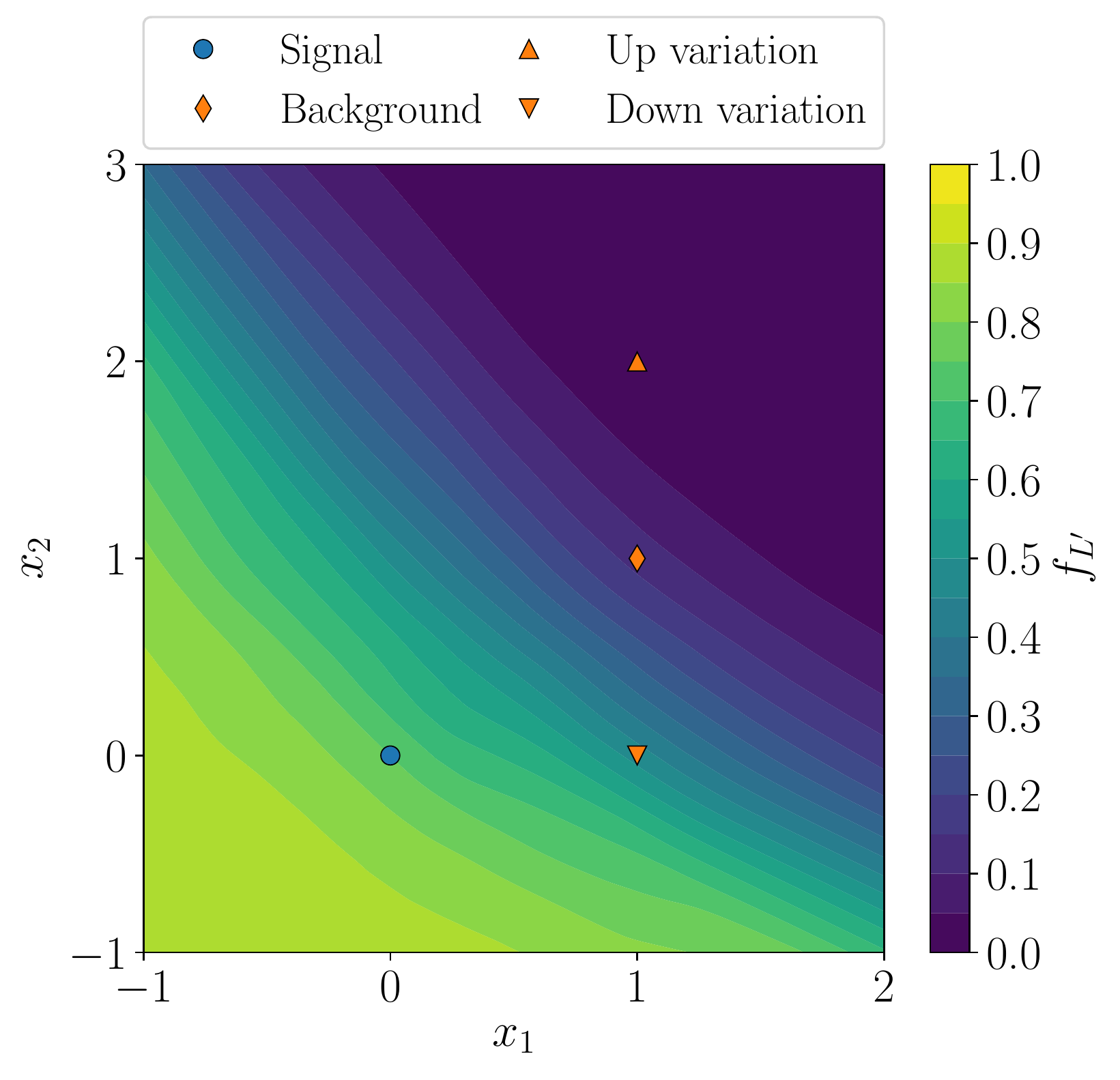}%
\includegraphics[width=0.45\linewidth]{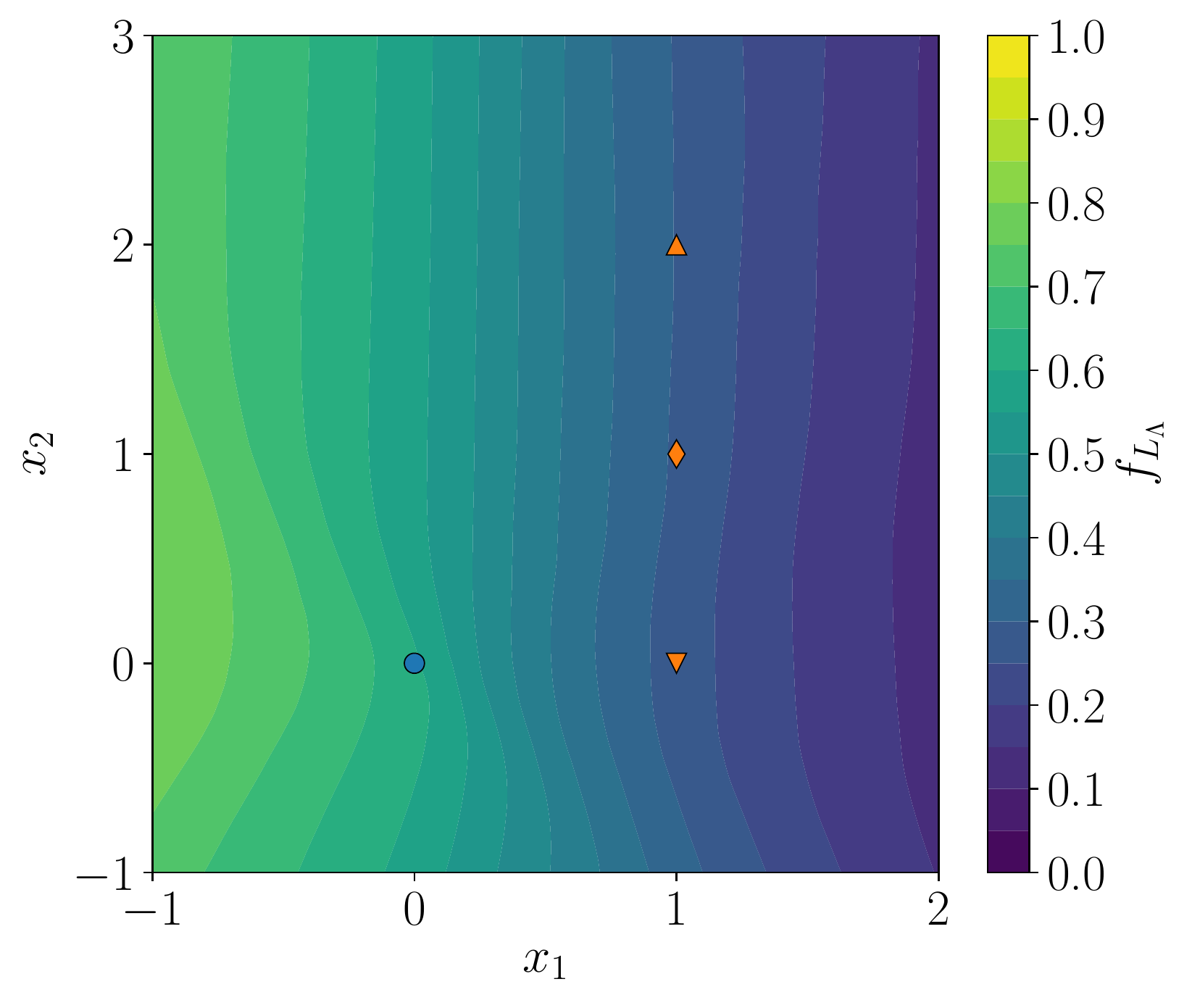}
\caption{The NN output as function of the input space, spanned by $x_{1}$ and $x_{2}$ (left) for the classifier trained with a cross-entropy function only ($f_{L^{\prime}}$), and (right) with an additional term penalizing the variation of the NN output with the systematic variation of $x_{2}$ ($f_{L_{\Lambda}}$). The markers indicate the mean values of the input distributions for the nominal and varied datasets.}
\label{fig:toy1_contour}
\end{figure*}

As can be seen from Fig.~\ref{fig:toy1_separation}, the approach successfully mitigates the dependence of the NN output on the variation of $x_{2}$ and therefore results in a classifier that is more robust in the presence of this systematic uncertainty. This is achieved on the expense of obliterating at least parts, if not all, separating information of $x_{2}$. Fig.~\ref{fig:toy1_contour} visualizes the NN output as a function of the input space spanned by $x_{1}$ and $x_{2}$. The additional penalty term, $\Lambda$, leads to the intended alignment of the surface of the NN output with the variation of $x_{2}$, resulting in similar values of the NN output for all realisations of the systematic variation. We find our approach to have an effect similar to the setup described in~\cite{louppe2017learning}.

\section{Application to a more complex analysis task typical for high-energy particle physics}
\label{sec:Application_to_a_more_complex}

In the following, we apply the proposed method to a more complex task typical for high-energy par\-ti\-cle phy\-sics. We use a dataset that has been released for the Higgs boson machine learning challenge described in Ref.~\cite{adambourdarios:hal-01208587}. This challenge uses a simplified synthetic da\-ta\-set from simulated collisions of high-energy proton beams with underlying hypothesized signal and background processes at the CERN LHC. The original target of the challenge was to separate events containing the decay of a Higgs boson into two tau leptons (signal) from all other events (background), to serve as benchmark for the success of different machine learning algorithms. The consideration of uncertainties, as required for a complete analysis of the data was not part of it. The dataset contains 30 input parameters, whose exact physical meanings are given in Ref.~\cite{adambourdarios:hal-01208587}. We split the dataset and use one third for training and validation of the NN and two thirds for deriving the following results.

\begin{figure*}
\centering
\includegraphics[width=0.4\linewidth]{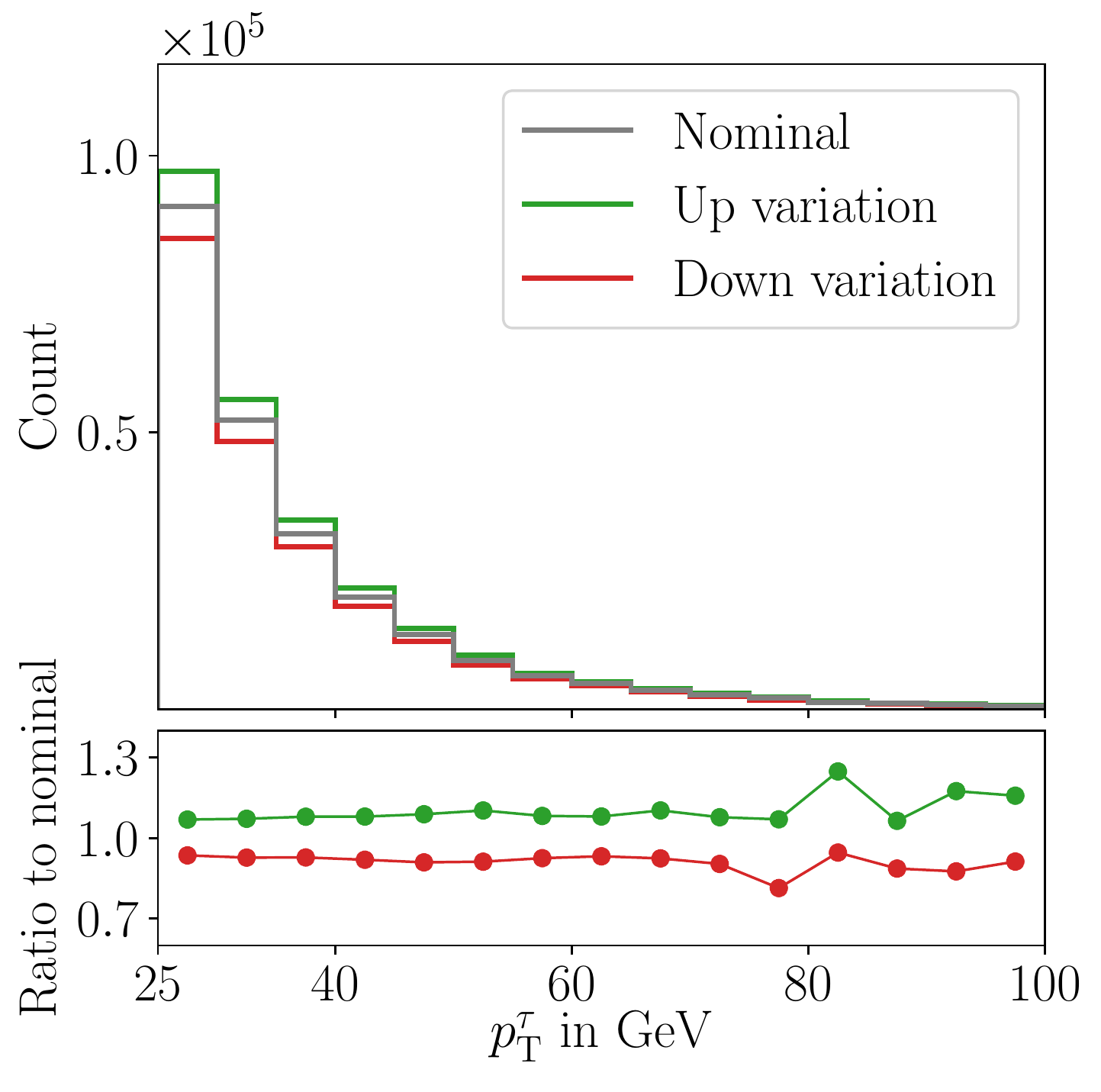}
\includegraphics[width=0.4\linewidth]{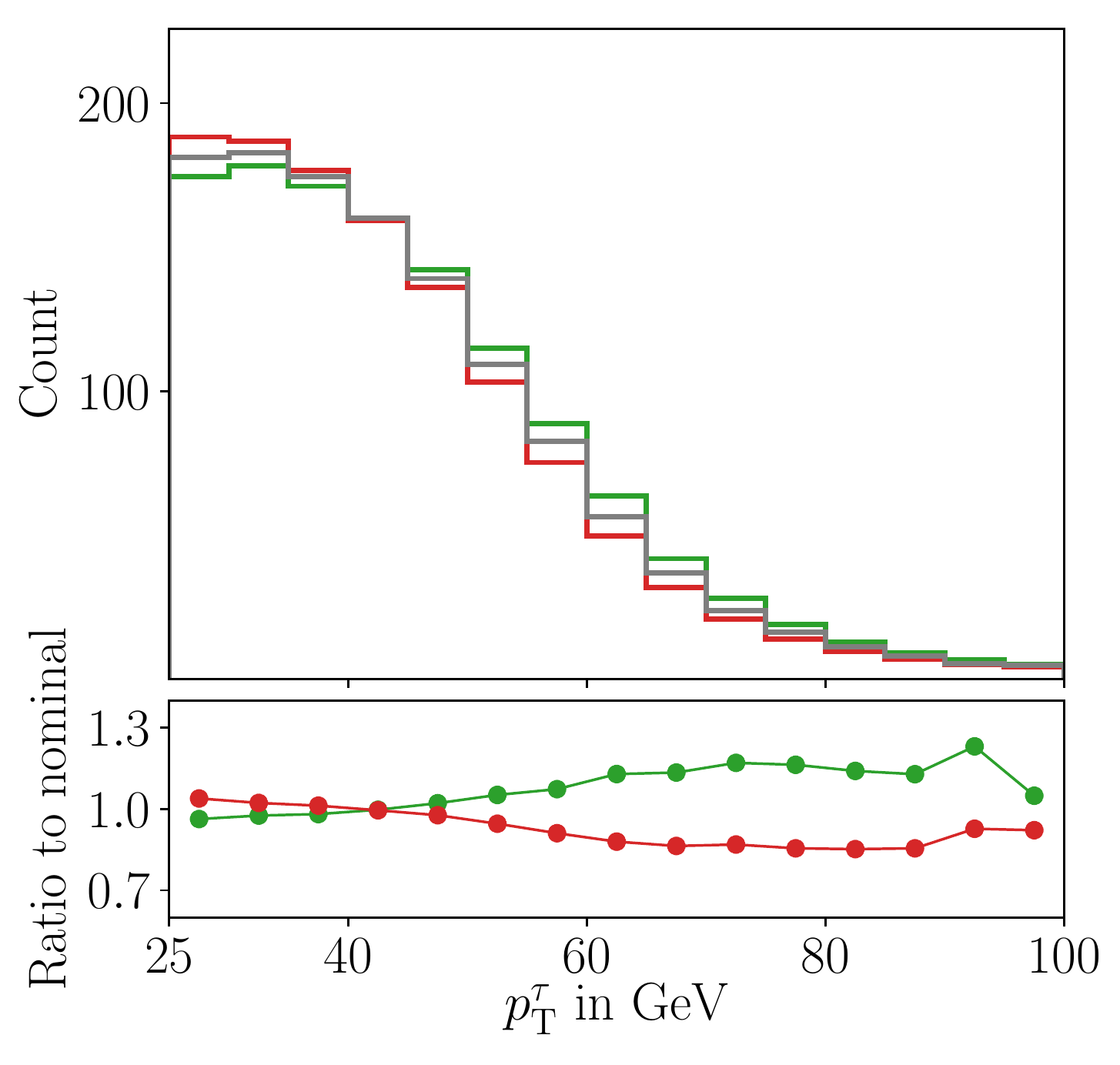}
\includegraphics[width=0.4\linewidth]{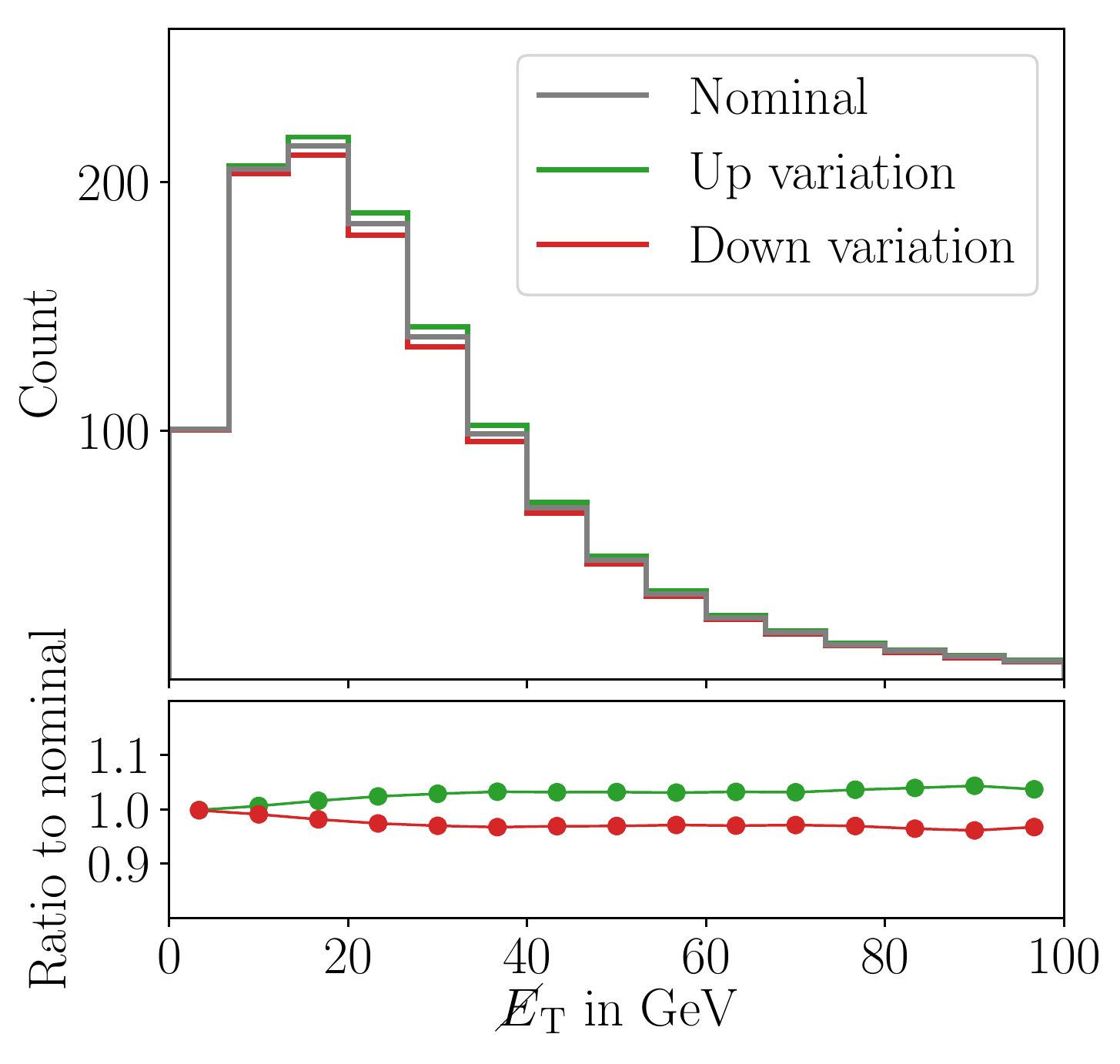}
\includegraphics[width=0.4\linewidth]{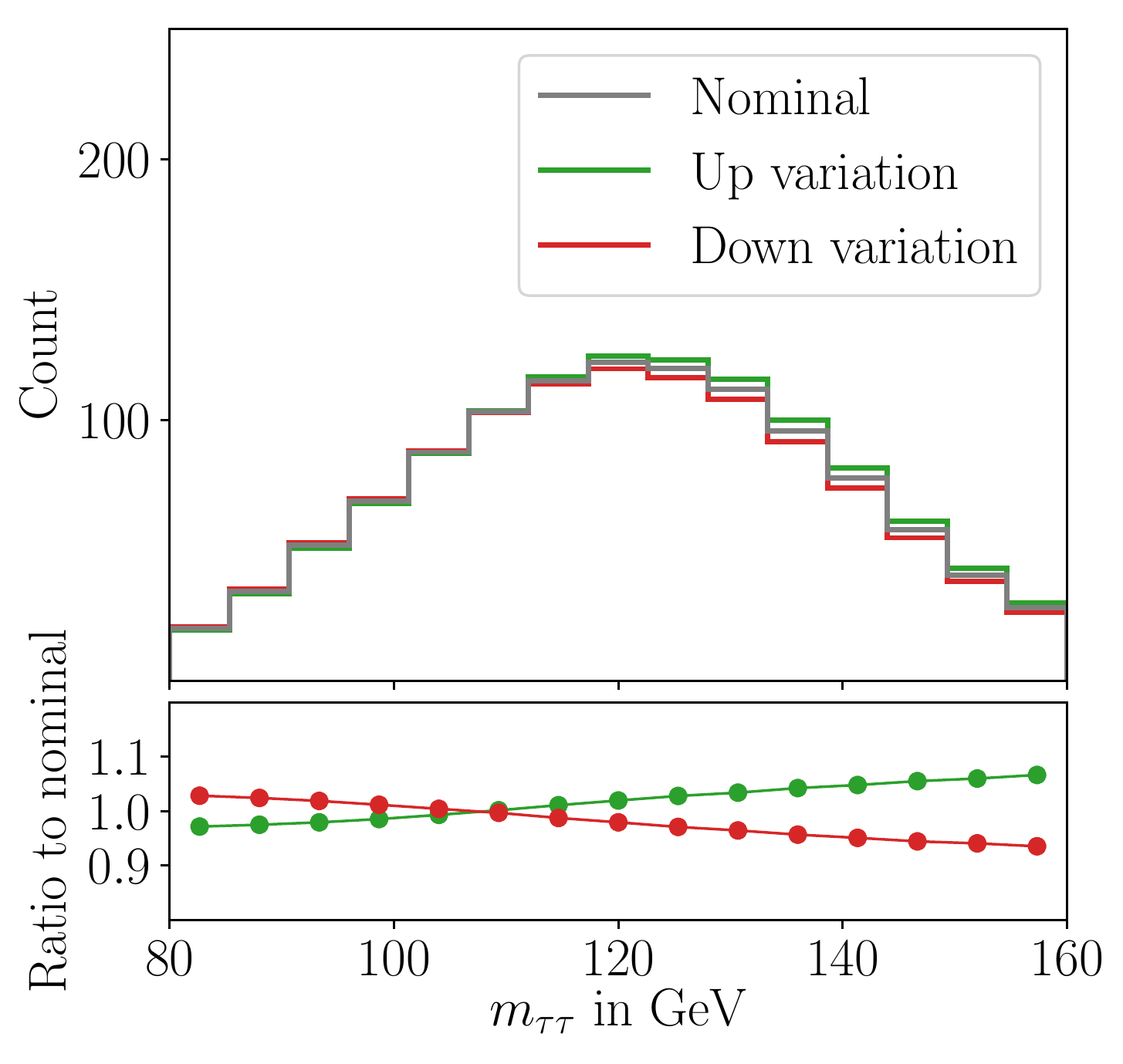}
\caption{Distribution of the transverse momentum of the hadronic $\tau$ decay $p_{t}^{\tau}$ (\texttt{PRI\_tau\_pt} in~\cite{adambourdarios:hal-01208587}), for the (upper left) background and (upper right) signal sample. Variations of this input parameter are introduced in form of statistical weights, i.e., for the $\pm3\%$ variation, subsamples with high (low) values of $p_{t}^{\tau}$ enter the analysis with a lower (higher) statistical weight than for the nominal sample. The weights for the background and signal sample, can be read off from the lower panels of these figures. Also shown are the (lower left) reconstructed missing transverse momentum and (lower right) the invariant di-$\tau$ mass estimated from the selected $\tau$ candidates, as described in Ref.~\cite{adambourdarios:hal-01208587}, from the signal sample, demonstrating the effect of the reweighting on variables correlated to $p_{t}^{\tau}$.}
\label{fig:higgs_taupt_control}
\end{figure*}

For our example, we use all parameters as input for the NN training. In addition, we introduce a systematic uncertainty, resembling the fact that the momentum and energy of a particle are the results of external measurements with a finite resolution. For our study we assume an uncertainty of $\pm3\%$~\cite{Aaboud:2018pen} on the transverse momentum of the reconstructed hadronic $\tau$ decay $p_{t}^{\tau}$, measured in GeV and labeled as \texttt{PRI\_tau\_pt} in Ref.~\cite{adambourdarios:hal-01208587}. The distributions of the nominal and varied input parameters are visualized in Fig.~\ref{fig:higgs_taupt_control} (upper row). To allow for migrations in and out of the selected input space due to the systematic variation we restrict the originally available dataset by raising the lower $p_{t}^{\tau}$ requirement from 20 to 25 GeV. For the background the distribution of $p_{t}^{\tau}$ is steeply falling. Thus the variation is dominated by migration effects at the lower $p_{t}^{\tau}$ boundary, resulting in an overall normalization uncertainty. The signal shows a maximum around $p_{t}^{\tau}\approx25\,\text{GeV}$, leading to a more apparent additional variation of the shape of the $p_{t}^{\tau}$ distribution, as shown in Fig.~\ref{fig:higgs_taupt_control} upper right. The dataset used for these results contains 814.9 (163750) weighted (unweighted) signal events and 162705.0 (238778) weighted (unweighted) background events using an additional scaling of the weighted number of signal events by a factor of two.

Instead of resampling the signal and background datasets with the varied values of $p_{t}^{\tau}$, we introduce the systematic variation in form of statistical weights. In this way we give a higher (lower) statistical weight to subsamples with low (high) values of $p_{t}^{\tau}$ with respect to the nominal sample. These weights are determined from the $p_{t}^{\tau}$ distributions shown in Fig.~\ref{fig:higgs_taupt_control} (upper row) for the background and signal sample, respectively. By construction all correlations across features of the input space are conserved by the reweighting, thus that reweighting $p_{t}^{\tau}$ leads to shape variations also of correlated observables, e.g., like the reconstructed missing transverse momentum or the estimate of the invariant di-$\tau$ mass, described in Ref.~\cite{adambourdarios:hal-01208587}, as shown in Fig.~\ref{fig:higgs_taupt_control} (lower row). We would like to emphasize that this reweighting technique is in fact the only way to apply a systematic variation of $p_{t}^{\tau}$ that respects the correlations to all other features of the input space on the given dataset. In a realistic analysis the reweighting technique is not meant to replace the resampling, but rather to complement it. A resampling could and should be applied, where correlations across input features may not be desired. To give an example, $p_{t}^{\tau}$ is mostly determined from track information. Therefore an uncertainty in the missing transverse momentum due to uniformity uncertainties in the calibration of the hadronic calorimeter should not impact $p_{t}^{\tau}$ with a correlation of 100\%. As in the case of the simple example of Section~\ref{sec:Application_to_a_toy} we use only the two discrete shapes corresponding to the $\pm3\%$ shifts in $p_{t}^{\tau}$, which are a sufficient input for the minimization of the loss function during the training process. Samples of intermediate realizations of these shifts have been checked to lead to the same result despite the more complex setup.

The NN has the same architecture as described in Section~\ref{sec:Application_to_a_toy}. For the implementation of $f_{L_{\Lambda}}$ we chose 20 equidistant bins in the range of $[0, 1]$ of the NN output for $\Lambda$, and $\lambda=20$. The batch size is set to $10^3$. The optimization of the trainable parameters is performed on $\SI{75}{\percent}$ of the training dataset and stopped if the loss has not decreased for 10 epochs in sequence, on the remaining part of the training dataset. The results are shown on an independent test dataset. We would like to emphasize that $\lambda=20$, is a free choice that has been made for illustrative purposes only. In a realistic application the optimal choice of $\lambda$ should be studied on a case by case basis.

\begin{figure*}
\centering
\includegraphics[width=0.4\linewidth]{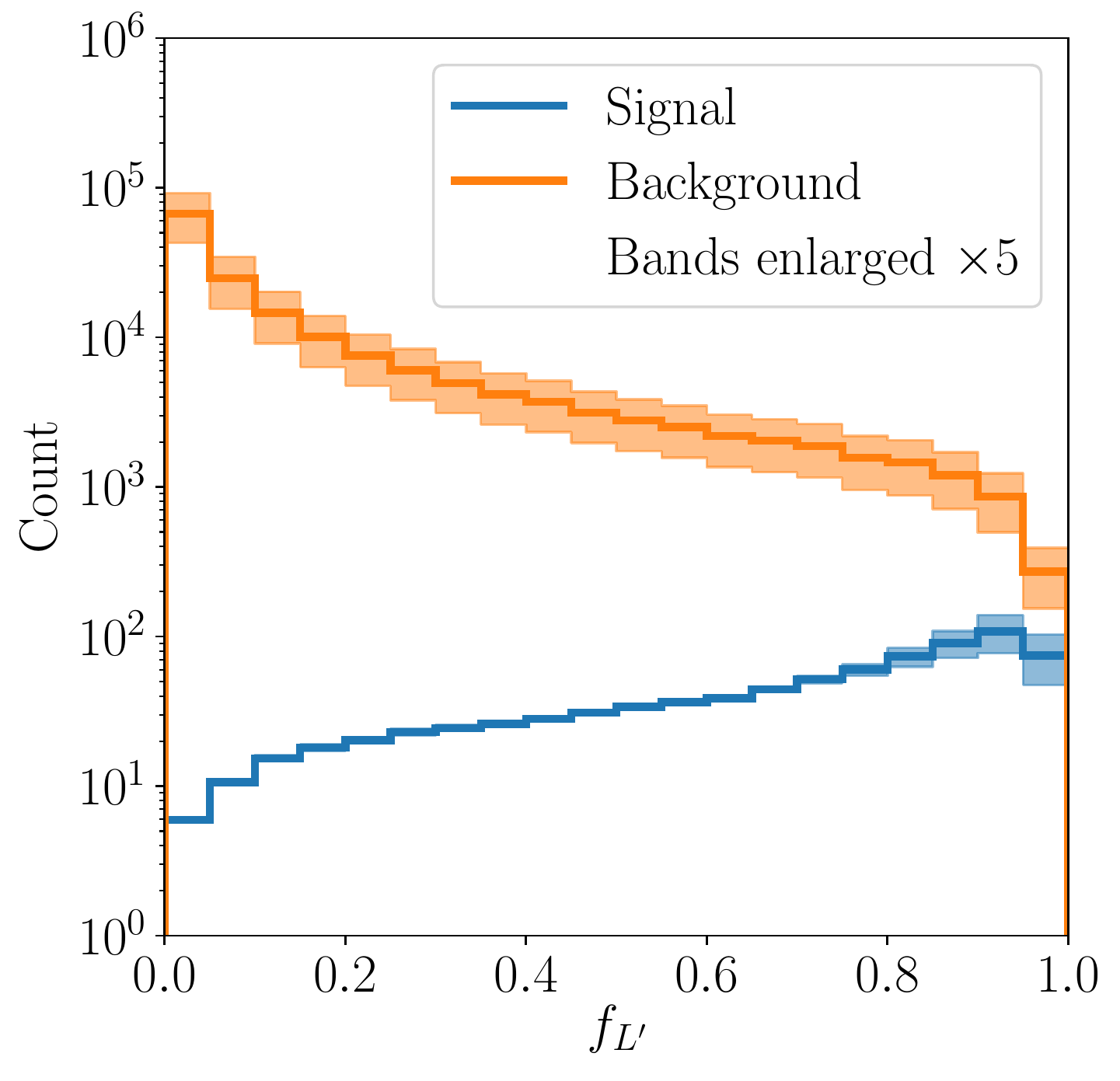}%
\includegraphics[width=0.4\linewidth]{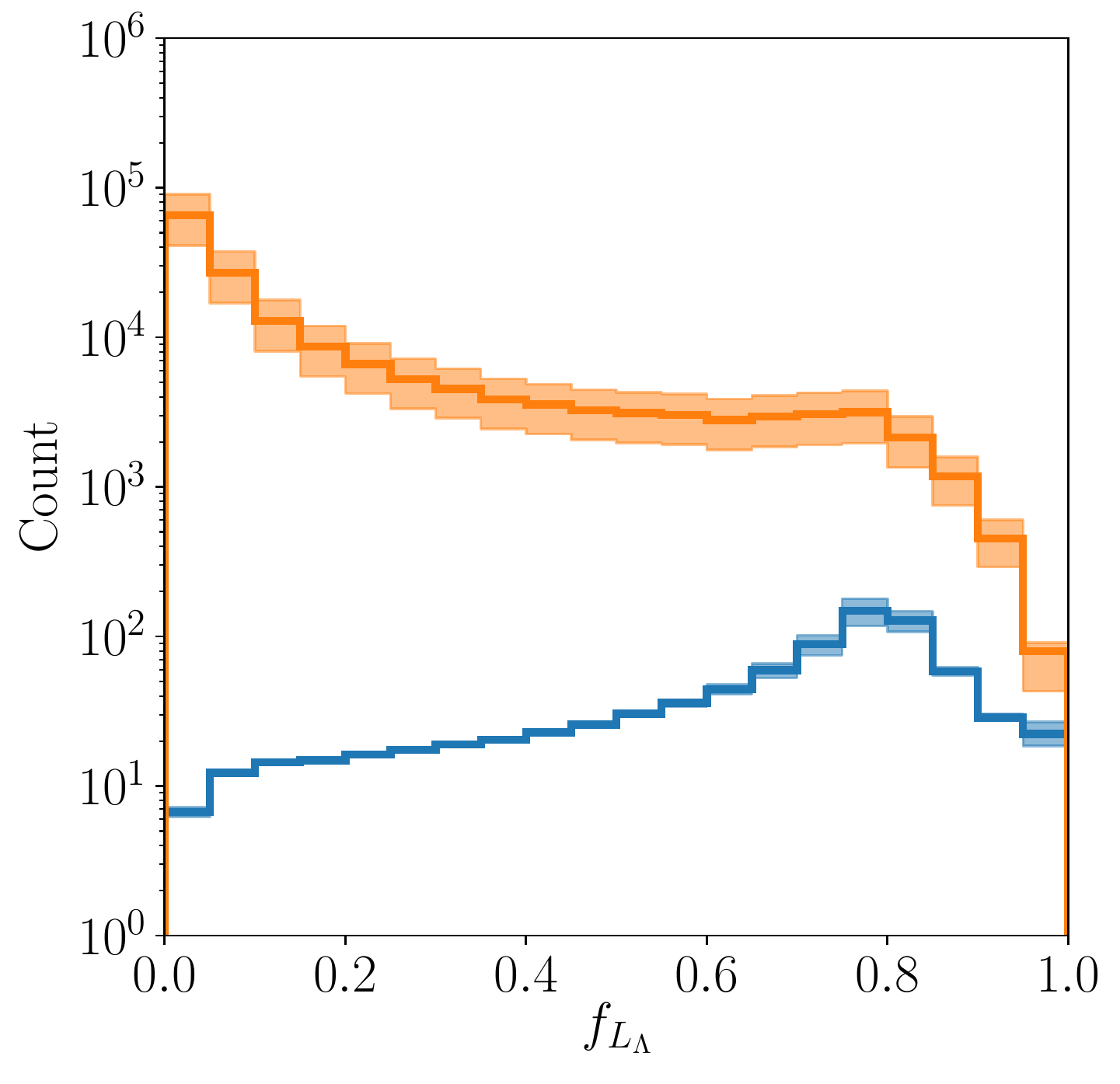}
\caption{
Distribution of the NN output for a classifier trained (left) with a cross-entropy function only ($f_{L^{\prime}}$), and (right) with an additional term penalizing the variation of the NN output with the systematic variation ($f_{L_{\Lambda}}$). The colored bands around the distribution of the NN outputs of the signal and background samples indicate the effect of the systematic variation of $(1.0 \pm 0.03)\,p_{t}^{\tau}$. For better visibility, the plotted bands are enlarged by a factor of five in both subfigures.}
\label{fig:higgs_separation}
\end{figure*}

In Fig.~\ref{fig:higgs_separation} the NN outputs $f_{L^{\prime}}$ and $f_{L_{\Lambda}}$ are shown. As in the case of the simple example given in Section~\ref{sec:Application_to_a_toy}, though less pronounced, the training based on a loss function including $\Lambda$ leads to a mitigated dependence of the NN output on the systematic variation of $p_{t}^{\tau}$. An important difference between both examples is that the uncertainty of the simple example given in Section~\ref{sec:Application_to_a_toy} is exclusively shape altering. In contrast to this the uncertainty variation in this more complex example includes a significant component acting on the normalization of the NN output, especially for the background distribution. A pure normalization uncertainty that does not lead to noticeable differences in the input space that can be related to its systematic variation can not be mitigated. In consequence a dominant overall normalization uncertainty, visible especially for the background distribution of $f_{L^{\prime}}$, is not significantly reduced by the use of $f_{L_{\Lambda}}$.

\begin{figure}
\centering
\includegraphics[width=0.8\linewidth]{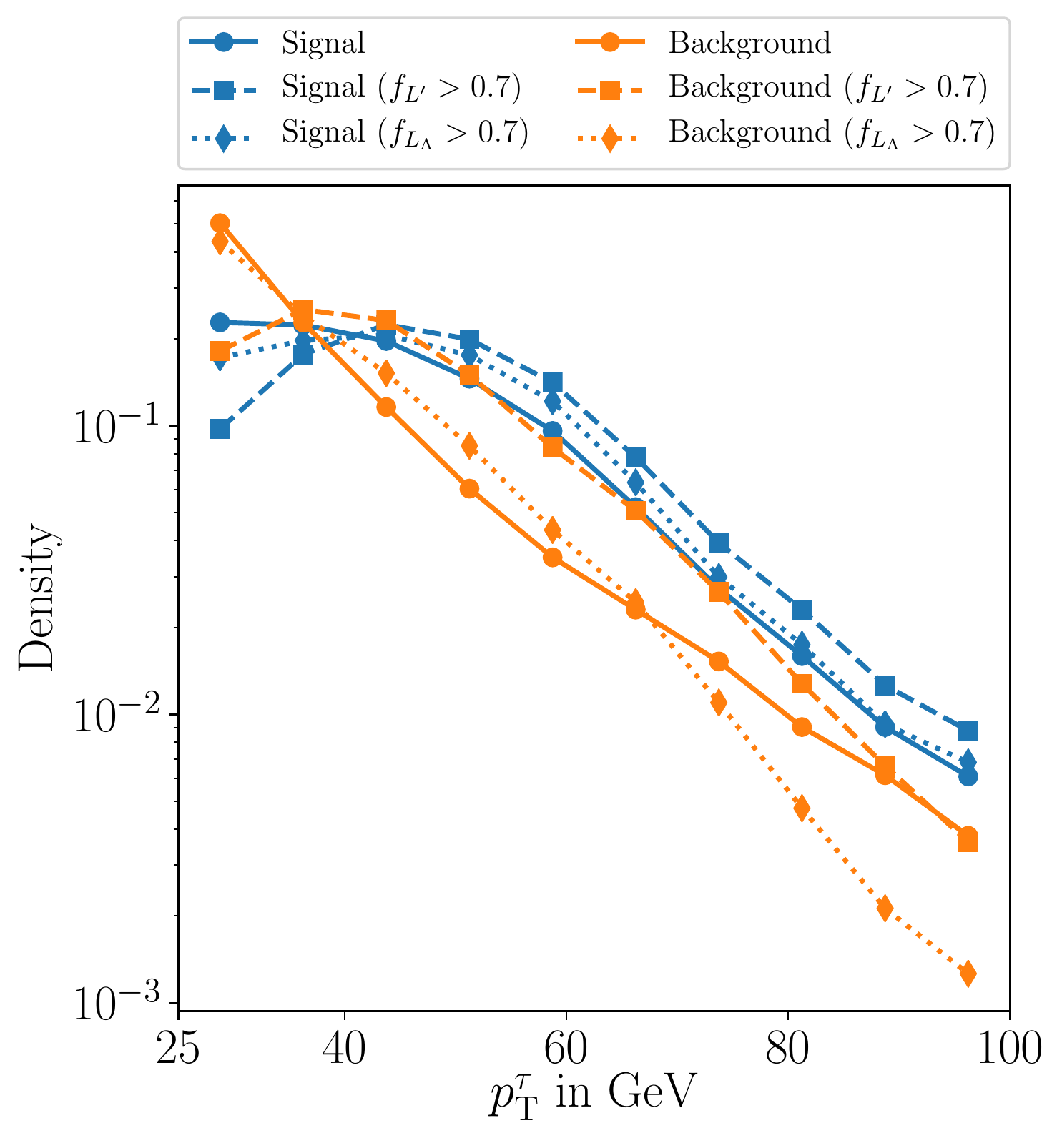}
\caption{Distribution of the transverse momentum of the hadronic $\tau$ decay $p_{t}^{\tau}$ (\texttt{PRI\_tau\_pt} in~\cite{adambourdarios:hal-01208587}). The distributions for signal and background are shown on the full unbiased sample, and two signal-enriched subsamples with $f_{L^{\prime}}>0.7$ and $f_{L_{\Lambda}}>0.7$.}
\label{fig:higgs_taupt}
\end{figure}

In Fig.~\ref{fig:higgs_taupt} the $p_{t}^{\tau}$ distributions for signal and background for the full sample, and for two signal-enriched subsamples are shown. The latter are obtained by a restriction of $f_{L^{\prime}}$ and $f_{L_{\Lambda}}$ to a value larger than $0.7$. On the full sample a generally harder $p_{t}^{\tau}$ spectrum for the signal is observed with a maximum around 45 GeV, in contrast to a steadily falling and softer spectrum for the background. In the signal-enriched subsample based on $f_{L^{\prime}}>0.7$ the $p_{t}^{\tau}$ distribution for the background is biased towards the same distribution as for signal. In the signal-enriched subsample based on $f_{L_{\Lambda}}>0.7$ this bias is alleviated and the $p_{t}^{\tau}$ distributions for signal and background are qualitatively unchanged with respect to the full sample.

At the LHC experiments the presence of the Higgs boson signal has been inferred from hypothesis tests based on a likelihood ratio between the case of including the Higgs boson signal and that of the null hypothesis without Higgs boson signal~\cite{atlas2011procedure}. Systematic uncertainties have been incorporated in form of nuisance parameters, which might be correlated, e.g., across processes, into the likelihoods. Best estimates and constraints on these nuisance parameters have been obtained by nuisance parameter optimization. The presence of the signal has been quantified, e.g., by means of its statistical significance in terms of Gaussian standard deviations (s.d.), in the limit of large numbers. To serve our discussion we emulate this discovery scenario, in a simplified way, constructing binned likelihoods for the signal and null hypotheses based on the histograms shown in Fig.~\ref{fig:higgs_separation}. In addition to the statistical uncertainties of the pseudo-data we incorporate the uncertainty indicated by the bands in Fig.~\ref{fig:higgs_separation} as process- and bin-correlated variations in the likelihoods, bound to a single nuisance parameter $\theta$, following the prescriptions of~\cite{atlas2011procedure}. The fit of a Higgs boson signal hypothesis with a single signal strength parameter of interest, $\mu$, to the pseudo-data, including the signal as expected by theory, leads to a constraint of the uncertainty in $\theta$ to 3\% of its initial value, both in the case of $f_{L^{\prime}}$ and $f_{L_{\Lambda}}$ as input distributions to the fit. This constraint is dominated by the power of the pseudo-data to determine the normalization related to $\theta$, especially in the first bins of the background dominated pseudo-data sample distribution, e.g., with more than 65 thousand counts in the first bin. When splitting the uncertainty into two independent nuisance parameters, $\theta_{\text{norm}}$ to govern the pure normalization uncertainty, and $\theta_{\text{shape}}$ to govern the pure shape altering uncertainty, we find the initial normalization uncertainty to be $7.6\%$ ($2.2\%$) for the background (signal) sample. We anticipate that the implementation with two independent nuisance parameters is not fully correct, but keeping this caveat in mind the study still serves the test we are interested in. After the fit of the Higgs signal hypothesis to the pseudo-data we observe the same constraint as on the uncertainty in $\theta$ before on the uncertainty in $\theta_{\text{norm}}$. We observe an ${\approx}35\%$ correlation between $\theta_{\text{norm}}$ and $\mu$. The constraint on the uncertainty in $\theta_{\text{shape}}$ is 0.8 (0.4) for $f_{L^{\prime}}$ and $f_{L_{\Lambda}}$ as input distributions to the fit, with a correlation of 55\% (5\%) to $\mu$. We observe similar results when performing a fit of the null hypothesis. The reduction of the correlation of $\theta_{\text{shape}}$ with $\mu$, when using $f_{L_{\Lambda}}$ instead of $f_{L^{\prime}}$ gives a quantitative measure in this case of the decorrelation of the shape altering part of the uncertainty with the parameter of interest.

\begin{figure}
\centering
\includegraphics[width=0.8\linewidth]{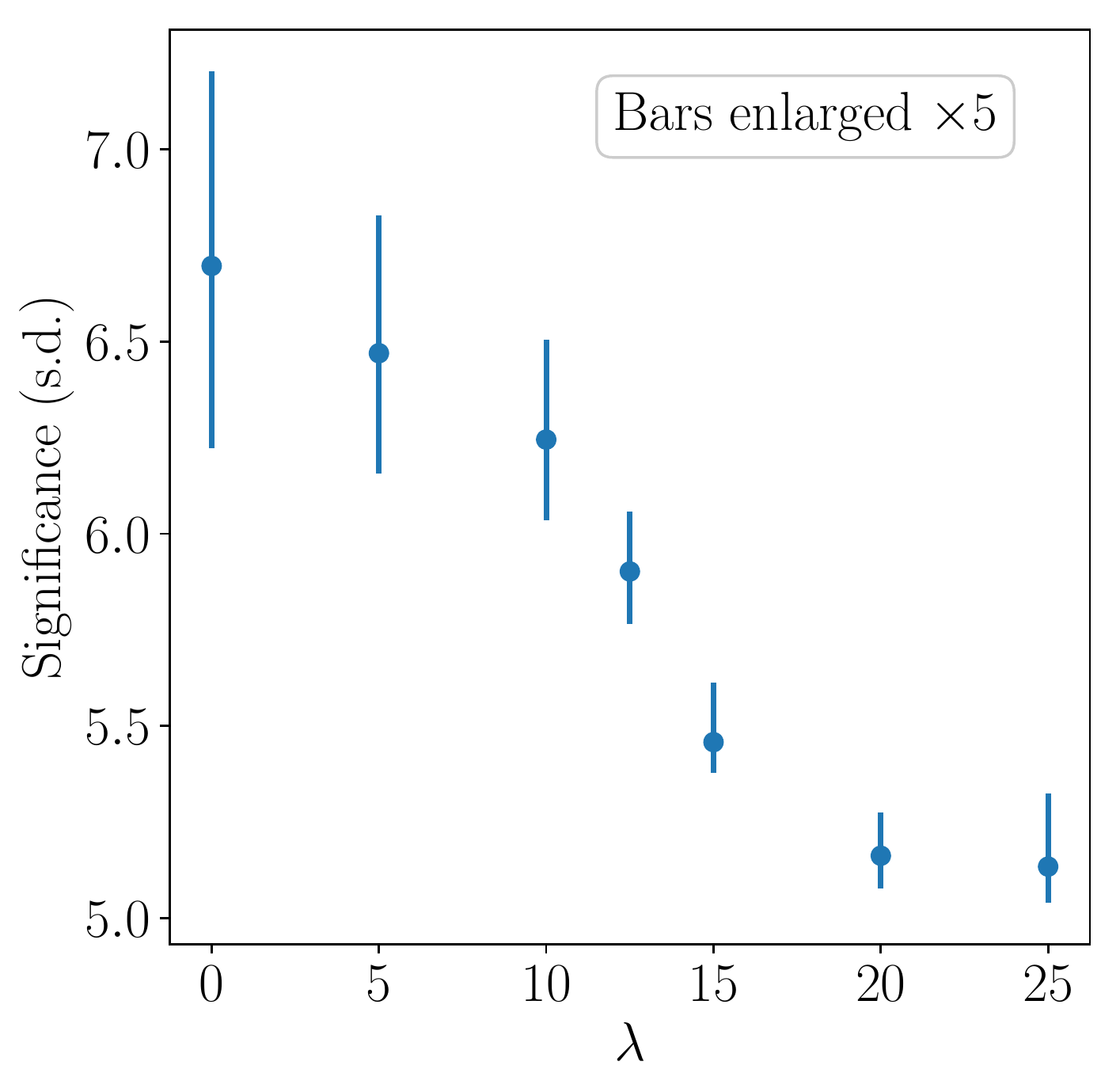}
\caption{Statistical significance of the Higgs boson signal in the dataset given in Ref.~\cite{adambourdarios:hal-01208587}, in standard deviations (s.d.), as a function of the tunable hyperparameter $\lambda$. The parameter value $\lambda=0$ corresponds to the choice of the distribution of $f_{L^{\prime}}$ as input to the fit to the pseudo-data. Increasing $\lambda$ further than shown here leads to another drop after $\lambda\approx 35$ approaching the significance for a single counting experiment that does not distinguish between signal and background, in the limit of $\lambda\to\infty$.}
\label{fig:higgs_shifts}
\end{figure}

In Fig.~\ref{fig:higgs_shifts} the significance of the analyzed signal in the pseudo-data, based on the fit to the null hypothesis is shown as a function of the hyperparameter $\lambda$, where $\lambda=0$ corresponds to $f_{L^{\prime}}$ as input to the fit. Using $f_{L^{\prime}}$ as input to the fit leads to a significance of 6.7 s.d., corresponding to a combined systematic and statistical relative uncertainty in the parameter of interest of $\Delta\mu/\mu=15\%$. This significance drops to a value of 5.2 s.d., corresponding to $\Delta\mu/\mu=19\%$, for $\lambda=20$. Such a drop is expected, since $p_{t}^{\tau}$ plays an important role in the separation of signal and background, not only as a single feature, but also via its correlations to other features in the input space~\cite{Wunsch:2018oxb}. The scan of $\lambda$ in this way visualizes to what extend the separation relevant information related to $p_{t}^{\tau}$ in the input space that is vulnerable to the variation of $p_{t}^{\tau}$, is masked during the training process for increasing values of $\lambda$. The information loss seems small for values of $\lambda\leq 5$ with a significant drop around $\lambda\approx10$ and a plateau around $\lambda\approx20$, which is the value we have chosen for our study. At this point most of the separation relevant information related to $p_{t}^{\tau}$ that is vulnerable to the variation of $p_{t}^{\tau}$ seems to be masked out from the training, such that $f_{L_{\Lambda}}$ turns mostly blind for $p_{t}^{\tau}$. Implicitly this can also be inferred from Fig.~\ref{fig:higgs_taupt}, where the distribution of $p_{t}^{\tau}$ qualitatively is the same for the signal-enriched and the inclusive samples.

In turn the uncertainty on the significance due to the systematic variation drops, roughly proportional to the loss in significance, from ${\approx}7.5\%$ (for $\lambda=0$) to ${\approx}1.8\%$ (for $\lambda=20$). We estimate the contribution of the systematic variation in $p_{t}^{\tau}$ to $\Delta\mu/\mu$, with $6.6\%$ (for $\lambda=0$), dropping to $1.8\%$ (for $\lambda=20$). At the same time, and with a larger slope, the absolute contribution of the statistical uncertainty to $\Delta\mu/\mu$ increases from $13.4\%$ (for $\lambda=0$) to $19.3\%$ (for $\lambda=20$), resulting in the overall decrease of the significance for increasing values of $\lambda$, for the given example. The loss in statistical power stems from the worse separation of signal and background for increasing values of $\lambda$, as also visible from Fig.~\ref{fig:higgs_separation}.

Increasing $\lambda$ to larger and larger values leads to another drop of the significance, which converges to the value for a single counting experiment that does not distinguish between signal and background, in the limit of $\lambda\to\infty$. This can be understood in terms of $\Lambda$ completely dominating the loss function thus that $L^{\prime}$ will more and more loose influence in the training task. As a consequence the NN will primarily be optimized on the suppression of the variation of $p_{t}^{\tau}$ rather than the separation of signal and background.

We would like to point out at the end of this discussion that it is usual practice in a measurement scenario to accept the increase of statistical uncertainty, which can in principle be controlled by an increase of the dataset for the benefit of a reduced sensitivity of the measurement on systematic variations of its input parameters, which might be difficult to control. We anticipate though that in the given scenario $\lambda=0$ remains the choice that maximizes the significance of the analysis despite its larger sensitivity to the systematic variation in this case. Our choice of $\lambda=20$ should be viewed as a free while still sensible choice to showcase the reduction of the influence of the systematic variation on the NN output.

\section{Summary}
\label{sec:Summary}

We have presented a new approach to reduce the dependence of the NN output to variations of features $x_{i}$ of the NN input space due to systematic uncertainties in the measured input parameters. We achieve this reduction by including the variation of the NN output w.r.t. the nominal value of $x_{i}$ in the loss function used for training. Compared to a previously published method of using an adversarial technique, the complexity of the presented method is reduced to one additional term in the loss function with less hyperparameters and no further trainable parameters. Systematic variations can be inscribed in the form of statistical weights, implying no further needs of reprocessing, further reducing the complexity of the training. Additional uncertainties just add to the sum of penalty terms in the loss function. In turn the method requires batch sizes large enough to populate the blurred histogram of the NN output used for the evaluation of the variation w.r.t the nominal value of $x_{i}$ in the loss function.

We have demonstrated the new approach with a simple example directly comparable to a solution of the same task exploiting the adversarial technique, and a more complex analysis task typical for high-energy particle physics experiments. In all cases the dependence of the NN output on the variation of a chosen input parameter is successfully mitigated. In application to a high-energy particle physics measurement this leads to a result less prone to systematic uncertainties, which is of increasing interest in the presence of growing datasets, where statistical uncertainties play a subdominant role in the measurement.

%\section*{Conflict of interest}
%
%On behalf of all authors, the corresponding author states that there is no conflict of interest.

\newpage
\bibliographystyle{splncs}
\bibliography{citations.bib}

\end{document}